\documentclass{aa}

\input{psfig.tex}
\begin{document}

\thesaurus{Section 3: Extragalactic astronomy
              (13.25.2;   X-rays: galaxies,
               11.01.2;   Galaxies: Seyfert,
               11.09.1 IRAS 09104+4109)}  

\authorrunning{A. Franceschini et al.: BeppoSAX uncovers a type-2 QSO in IRAS 09104+4109
}
\title{BeppoSAX uncovers a type-2 QSO in the 
hyperluminous infrared galaxy IRAS 09104+4109}

\offprints{A. Franceschini}

\author{A. Franceschini\inst{1}, L. Bassani\inst{2}, M. Cappi\inst{2}, 
G.L. Granato\inst{3}, G. Malaguti\inst{2}, E. Palazzi\inst{2}, M. Persic\inst{4}}

\institute{$^1$ Dipartimento di Astronomia, Vicolo Osservatorio 5, I-35122 Padova,
Italy; E-mail: franceschini@pd.astro.it \\
$^2$ ITeSRE/CNR, via Gobetti 101, I-40129 Bologna, Italy\\
$^3$ Osservatorio Astronomico, Vicolo Osservatorio 5, I-35122 Padova, Italy\\
$^4$ Osservatorio Astronomico, Via G. Tiepolo 11, I-34131 Trieste, Italy}

\date{Received 15 July 1999/ Accepted 3 September 1999}

\titlerunning{A. Franceschini et al.: SAX uncovers a type-2 QSO in IRAS 09104+4109}
\authorrunning{A. Franceschini et al.: SAX uncovers a type-2 QSO in IRAS 09104+4109}

\maketitle

\begin{abstract}

We studied with BeppoSAX   the infrared luminous galaxy IRAS 09104+4109              
over a very wide X-ray band from 0.1 to 80 keV. Our   observations indicate
the dominance of a thermal component at energies below 8 keV, which we attribute 
to the free-free emission from the intracluster (IC) plasma surrounding the source. 
Above 10 keV we find evidence for
the existence of flux in excess with respect to the free-free 
IC plasma emission. This, together with the marginal detection of a neutral
iron line at $\sim 6.4$ keV, gives a strong indication for the presence of an AGN
deeply buried within the source. This component is best modelled by a strongly absorbed 
(N$_{\rm H}$ $\ge$ 5 $\times$ 10$^{24}$ cm$^{-2}$) power-law plus 
unabsorbed reflection spectrum (0.15 $\le$ R $\le$ 0.3),
which is also responsible for the cold iron line. The unabsorbed broad-band
(2-100 keV) X-ray emission of this AGN is 
2.5 $\times$ 10$^{46}$ erg s$^{-1}$, well within the 
range of quasar luminosities. Our results indicate
that IRAS 09104+4109 is indeed the prototype of a rare class
of sources, the luminous type-2 QSOs. The association of this source with
a huge cooling-flow of $\sim 1000\ M_\odot$ in the cluster, 
as indicated by the X-ray data, might suggest that such condition of extremely fast 
mass accumulation could favour the survival of a thick obscuring envelope, which
would otherwise be quickly destroyed by the very luminous central source.

\keywords{galaxies: individual:IRAS 09104+4109- galaxies: active-
X-rays: galaxies}
\end{abstract}

\section{Introduction}

The nature of Hyperluminous Infrared Galaxies (HyLIRGs, i.e. sources with bolometric 
luminosity in excess of 10$^{12}$ L$_\odot$, mostly emitted in the mid/far-IR) has been the
subject of a lively debate since their discovery by IRAS.
Observations have been accumulated in recent years, but, in spite of recent
significant progresses based on mid/far-IR spectroscopy (Genzel et al. 1998),
the origin of their high luminosity remains uncertain, with massive
starbursts and AGN both suspected to be responsible for the huge observed flux.
Although optical spectroscopic data often indicate the presence of an AGN in HyLIRGs,
several of them have not been detected (yet) in X-rays (Ogasaka et al. 1997), 
so the idea that they harbour a powerful QSO is still to be verified.
If the dominant radiation power is due to an AGN, this must be deeply buried within
large amounts of absorbing material because these sources appear as underluminous in
soft X-rays with respect to normal AGNs.
Hard X-ray measurements (above 10 keV) are then particularly effective in probing the 
presence of an optically-hidden AGN, as they penetrate column densities as high as 
several $\times$ 10$^{24}$ cm$^{-2}$.
Testing this would bear relevant implications for the long sought high redshift/luminosity 
analogues of Seyfert-2 galaxies, which are required by the
unified model of AGN, and for the origin of the X-ray and IR cosmic backgrounds.

Among HyLIRGs, certainly one of the most intriguing is IRAS 09104+4109 (z=0.442), 
which is the most luminous object in the Universe with $z<0.5$ and one of the most luminous 
IR-galaxies known (L$_{\rm FIR} \sim 2\times 10^{46}\ erg s^{-1}$). 
Kleinmann et al. (1988) have shown it to have a Seyfert2-like spectrum,
with strong narrow emission lines. Later Hines \& Wills (1993) and Hines et al.
(1999) reported the detection of broad MgII and Balmer emission lines  
in the highly polarized spectrum. Both near- and mid-IR
spectroscopy of the source provided evidence for the presence of a dusty torus
(Taniguchi et al. 1997, Evans et al. 1998); this was further confirmed
by HST imaging polarimetry of the source (Hines et al. 1999). 
 An ASCA observation of IRAS 09104+4109
revealed it as a powerful X-ray source, with a power law spectrum 
(photon index $\sim$ 2) similar to what is observed in most AGNs 
and a strong  line at 6.67 keV, interpreted as due to helium-like
iron scattered into our line of sight (Fabian et al. 1994a).
However, this is not an unequivocal interpretation of the data since the source
resides in the core of a rich cluster of galaxies, which also emits in X-rays.
Indeed a follow-up ROSAT HRI image 
of this sky region confirmed that the X-ray emission is extended and 
dominated by a
cooling flow (Fabian \& Crawford 1995); given the similarity of the
extrapolated ASCA flux to that found by ROSAT, it is likely that
the bulk of the 0.1-10 keV  emission is also dominated by radiation
from the hot gas in the surrounding cluster and not by the AGN.
In fact ASCA data were better remodelled by an isothermal gas and an 
absorbed cooling flow, although the 
strength of the iron line was such that a contribution from a hidden nucleus 
could not be excluded. 

In this paper we report X-ray observations with BeppoSAX of IRAS 09103+4109,
and in particular the detection of excess emission at energies greater than 10 keV, 
which is likely due to non-thermal quasar emission emerging from a thick 
absorbing torus, well in excess of the thermal emission from the cluster. We will argue that 
this may be the first high-luminosity type-2 QSO detected in X-rays.

In Sect. 2 we report about the observations and data reduction, while Sect. 3 expands
on the analysis of the spectral data. Our results for the fitting of the 0.1-10 keV
spectrum are reported in Sect. 4, and in Sect. 5 those for the higher energies.
The very complex nature of this source revealed by BeppoSAX is discussed in Sect. 6.
We assume $H_0=50$ km/s/Mpc throughout the paper.

\section{Observation and data reduction}

The BeppoSAX X-ray observatory (Boella et al. 1997a) is a major programme
of the Italian Space Agency with participation of the Netherlands Agency for 
Aereospace Programs. 
This work concerns results obtained with three of the 
Narrow Field Instruments (NFI) onboard:
the Low Energy Concentrator Spectrometer (LECS; Parmar et al. 1997), 
the Medium Energy Concentrator Spectrometers 
(MECS; Boella et al. 1997b), and with the Phoswich 
Detector System (PDS; Frontera et al. 1997).
LECS, MECS and PDS operative energy bands are 0.1--4.5 keV, 1.5--10 keV 
and  13--80 keV, respectively. 
The data from the HPGSPC did not provide significant count rate
and therefore will not be considered here.

BeppoSAX NFI pointed at IRAS P09104+4109 from Apr 18th, to Apr 19th, 1998.
The reduction procedures and screening criteria used to produce
the linearized and equalized (between the two MECS) event files
of the LECS and MECS were standard (Guainazzi et al. 1999). 
PDS data were analyzed
using the XAS reduction procedure which takes into account rise
time and spike corrections (Chiappetti \& Dal Fiume 1997).
The effective on-source exposure times were $2.42\times10^4$ s for the LECS,
$5.47\times10^4$ s for the MECS, and $3.43\times10^4$ s for the PDS.
Spectral data were extracted from regions centred on IRAS 09104+4109
with radii of 2$'$ and 4$'$, matching the PSF FWHM, for the MECS and the LECS respectively. 
LECS and MECS background subtraction was performed by means of blank sky spectra
extracted from the region around the source.

The PDS products were obtained by plain subtraction of the "off-" from
the "on-source" data. 
The net source count rates were (1.53$\pm$0.09)$\times10^{-2}$ c/s in the LECS
(0.1--4.5 keV), and (1.72$\pm$0.05)$\times10^{-2}$ c/s in 
the two MECS (2--10 keV).
The net source count rate in the PDS was 0.107$\pm$0.026 c/s
between 13 and 80 keV, which gives a detection at 3.3 $\sigma$
level also after a
conservative subtraction of the systematic residuals which
are currently evaluated at $\sim$0.02 c/s in the 13--200 keV band
(Guainazzi \& Matteuzzi 1997).

\section{Spectral analysis}

LECS and MECS data were rebinned in order to sample the energy resolution of the
detector with an accuracy proportional to the count rate: one channel
for LECS and 5 channels for MECS.
Spectral data from LECS, MECS and PDS have 
been fitted simultaneously. Normalization constants have been 
introduced to allow for known
differences in the absolute cross-calibration between the detectors. The values
of the two constants have been allowed to vary
and turned out to be within $\sim$5\% of the suggested values (see 
Fiore et al. 1999).
The spectral analysis has been performed by means of the {\sc XSPEC 10.0} package,
and using the instrument response matrices released by the BeppoSAX Science Data 
Centre in September 1997. 
All quoted errors correspond to 90\% confidence intervals  
(for one interesting parameter this corresponds to $\Delta\chi^2$ = 2.71).
Source plus background light curves did not indicate significant flux variability.
Therefore the data from the whole observation were summed together for the 
spectral analysis.

All the models used in what follows contain an additional term to allow for the
absorption of X-rays due to our Galaxy, that in the direction of IRAS P09104+4109 amounts
to $1.81\times10^{20}$ cm$^{-2}$ (Murphy et al. 1996). 
Given the high $z$ of the source, all the spectral parameters 
are given in the reference system of the emitting source unless
otherwise specified.

\section{The 0.1--10 keV spectrum}

To check the consistency
with previous X-ray data we first concentrate on the 0.1-10 keV band and 
fit the LECS-MECS data with an absorbed power law plus a
narrow ($\sigma\equiv$0) line to account for excess emission
around 6-7 keV, as done by Fabian et al. (1994a) for ASCA data.
The fit is satisfactory 
($\chi^{2}/\nu$=48.1/60) and results in a spectrum having a  
photon index   
$\Gamma$=1.9$\pm$0.1 and an absorbing column density 
N$_{\rm H}$=(1.4$^{\rm+1.8}_{\rm-0.9}$) $\times$ 10$^{21}$ cm$^{-2}$; 
the line is centered at 6.62$\pm$0.12 keV
and has a rest frame equivalent width (EW) of 
1026$^{\rm+365}_{\rm-253}$ eV. If the line width is allowed to
vary, the additional parameter gives only an 80$\%$ 
improvement in the fit ($\Delta \chi^{2}/\nu \simeq 3$ for one additional
parameter) with a rest frame width of (0.3$\pm0.2$ keV). 
These model parameters are in agreement with 
those determined by Fabian et al. (1994a) except for the line 
equivalent width, which is 
higher in the BeppoSAX observation compared with the ASCA one 
(444$^{\rm+120}_{\rm-173}$ eV), but still consistent at 2$\sigma$ level. 

The unabsorbed rest frame luminosity in the 0.5-10 keV band is similar
to the one determined by ASCA (1.9 $\times$ 10$^{45}$ erg s $^{-1}$), 
indicating that the emission did not change over years. 
In view of this lack of variability and of the ROSAT resolved images 
of a diffuse IC source at low energies (Fabian \& Crawford 1995),
a more appropriate description of the 0.1-10 keV spectrum  is in terms of
thermal emission from the hot gas in the cluster surrounding the source.
Consequently we have fitted the data both with a Bremsstrahulung
model plus a gaussian line to parameterize the Fe line energy and intensity
and with the XSPEC/MEKAL model to estimate the metal abundances
\footnote{MEKAL is a program of the XSPEC library modeling the emission 
of a hot IC plasma (including free-free continuum, edge and line emission), 
as a function of the plasma temperature kT, ion and electron density,
source redshift $z$, and average metal abundance Z, assumed solar abundance ratios.}.

A good fit is obtained  in the first case ($\chi^{2}/\nu$=47.3/61)
for a gas temperature of kT=9.7$^{\rm+2.7}_{\rm-1.8}$ keV and 
no absorption in excess of the Galactic value.
The line energy and equivalent width are consistent with the values found
with the power law model ($6.63\pm0.12$ keV, 
EW=920$^{\rm+362}_{\rm-237}$ eV).
Also in this case the line is compatible with zero width and so in the
subsequent analysis the line was always assumed to be narrow.

The MEKAL model gives a best-fit temperature of 8.3$^{\rm+1.9}_{\rm-1.5}$ keV
and a metal abundance consistent with solar, i.e $Z\simeq 1.2\pm 0.5\ Z_\odot$ 
($\chi^{2}/\nu$=51.9/62). This fit in count units and the corresponding
residuals are reported in Fig. 1.
According to the correlation found by Fabian et al. (1994b), however,
this abundance is unusually high for the measured temperature,
a better value being $Z\le$ 0.4 solar (in particular for a cluster at this large redshift), 
in which case
kT becomes $\ge$ 9 keV. Also note that some residual emission is present around the iron
line peak energy when using the MEKAL model (see Fig. 1). This explains 
the slightly better fit and lower line energy (compared with the
expected 6.7 keV) obtained with the Bremsstrahlung model.

\begin{figure}
\psfig{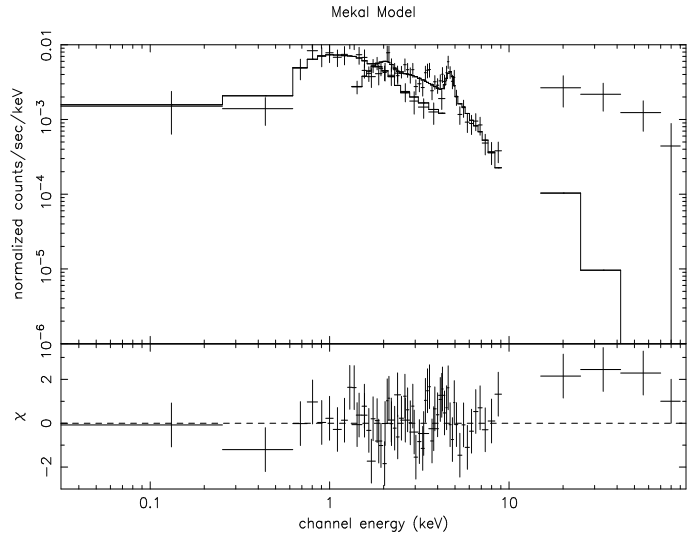}
\caption{BeppoSAX LECS, MECS and PDS data of IRAS 09104+4109 fitted with
a MEKAL model (top panel). The residuals beween the data and the model are plotted
at the bottom with error bars.}
\end{figure}

In view of these inconsistencies, we have added a narrow gaussian line
to the MEKAL model: this extra line turned out to be centered at 
6.1$^{\rm+0.4}_{\rm-0.3}$ keV (rest-frame) and characterized by a 
rest-frame EW=406$^{\rm+174}_{\rm-316}$ eV with respect to the thermal
continuum. The addition of this  extra line provides an
improvement of the fit ($\Delta\chi^{2}$=5.2 for two 
additional
parameters) which is significant at more than 90$\%$ confidence level.
Although the cold line is poorly constrained in the contour plot of the line energy 
versus normalization, it is however separated from the warm iron line at the 90$\%$
confidence level (see Fig. 2, where the line fitting is done using
the Bremsstrahlung model).

\begin{figure}
\hspace{1cm}
\psfig{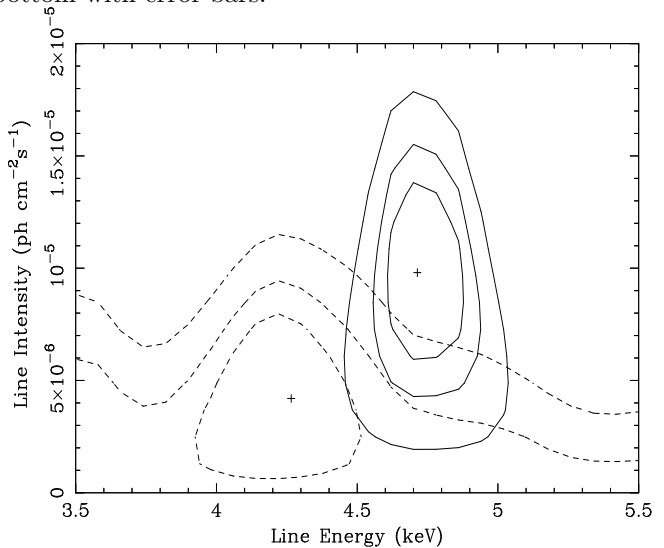}
\vspace{0.5cm}
\caption{Contour plot of line energy versus intensity for the 6.4 keV
line (dashed line) and 6.7 keV line (continuous line) in the
observer rest frame, assuming a Bremsstrahlung model.
The best-fit EW (observer rest frame) are 
78$^{\rm+204}_{\rm-58}$ eV and 493$^{\rm+397}_{\rm-302}$ eV for the
cold and warm iron line respectively. The three contour levels correspond
to 68\%, 90\% and 99\% rejection probability.
}
\end{figure}

Also note that when in the MEKAL model the iron abundance is fixed to 0.4,
i.e. a value more appropriate for the cluster temperature observed,
the 6.4 keV line becomes significant at more than 99$\%$ confidence and
has an equivalent width of 390$^{\rm+272}_{\rm-151}$ eV in the
quasar rest frame.

We have also tried to add an absorbed cooling flow to the MEKAL model
(as done by Fabian \& Crawford 1995 with the ASCA data) 
but found that this extra component was not required by the data:
the quality of the fit does not improve and the mass cooling rate
is poorly constrained ($\le$ 1520 M$_\odot$ yr$^{-1}$).

In conclusion, the LECS-MECS data basically confirm previous X-ray results
(Fabian et al. 1994a; Fabian \& Crawford 1995) and give a further indication
that the bulk of the 0.1 to 10 keV emission is dominated by the cluster thermal
component. We do not find strong evidence at these energies
neither for a cooling flow nor for quasar emission. However, the 
marginal detection of a neutral iron line already suggests the presence of an AGN, 
although providing a minor contribution to the continuum at these energies.

\section{The AGN nucleus unconvered by the PDS}

Extrapolation of the best fit MEKAL model discussed above fails to reproduce the
higher energy PDS data, indicating the existence of excess emission above 10 keV
(see Fig. 1). However, given the large field of view of the PDS (1.3$^\circ$
FWHM) care must be taken to check that no hard X-ray emitting sources could be
present in the target field and contaminate the PDS signal. Inspection of the
$\sim 45$ arcminutes diameter mapped by the SAX NFI reveals the presence of 5
marginal ($S/N>3$) sources in the 0.1-2 keV energy channel, none of which is
present in the highest energy map at 5-10 keV. So no high-energy sources, apart
from IRAS 09104+4109, are indicated in the inner PDS field. Further inspection of
the sky region surrounding IRAS 09104+4109 using other X-ray catalogs confirms
that a few low X-ray energy sources are contained within the PDS field of view.
Of the 4 RASS objects found, one corresponds to IRAS 09104+4109 (0.1213 c/s)
two are quasars at redshifts 0.936 (0.074 c/s) and 0.7325 (0.063 c/s) and one is
a normal galaxy (0.111 c/s) (Bade et al. 1998). At higher energies (2-10 keV) we
find a source in the HEAO-1 A1/A3 catalogues as well as in the ASCA-SIS database,
which correspond to the IRAS source. We therefore conclude that the IRAS source
is the dominant one in the 2-10 keV band and presumably also in the PDS energy
range.

In what follows we model the broad-band BeppoSAX spectrum
of the source, trying to reconcile the 0.1-10 keV data with the
excess emission observed at higher energies and with the possible presence
of a neutral iron line. 

\begin{figure}
\psfig{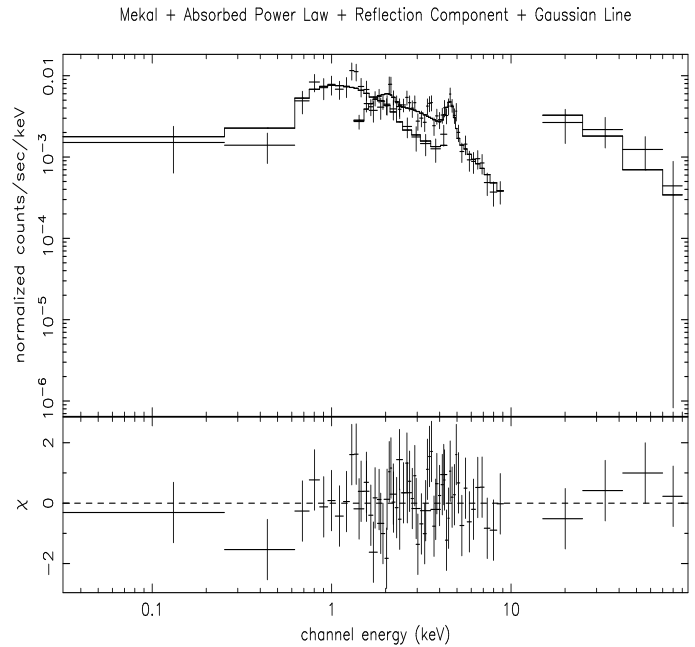}
\caption{Our best-fit model of the BeppoSAX X-ray spectrum of
IRAS 09104+4109, in counts units. The model includes a MEKAL component
(kT=5.5 keV and metal abundance $Z= 0.4\ Z_\odot$) plus an absorbed power law 
(N$_{\rm H}$=7 $\times$ 10$^{24}$ cm$^{-2}$ and photon index $\Gamma$=1.9 )  
with associated reflection component (reflected fraction $R=0.2$) and neutral iron line 
(observed EW=1.3 keV); the residuals beween the data and the model are plotted
in the bottom panel.}
\end{figure}

\begin{figure}
\psfig{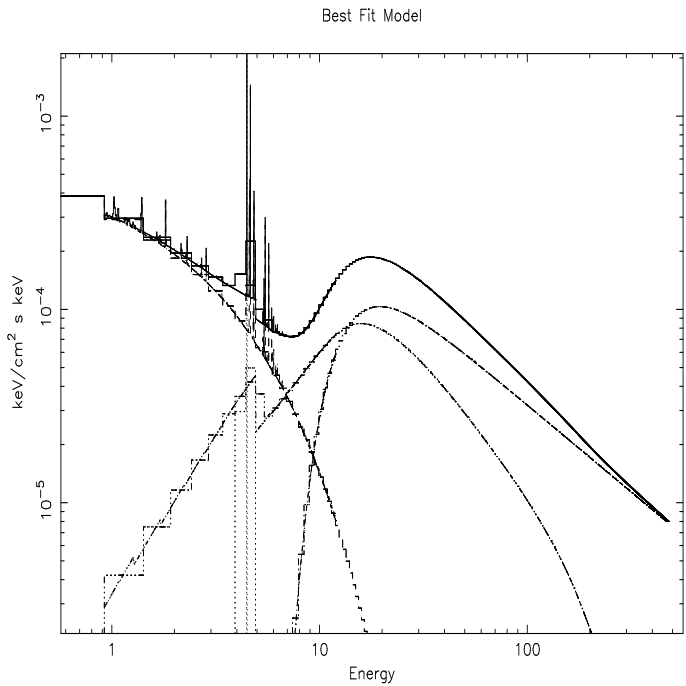}
\caption{Same best-fit model of the BeppoSAX X-ray spectrum of
IRAS 09104+4109 as in Figure 3, but in physical units. The model includes an IC plasma 
component (the line with exponential cutoff at 5 keV),
plus an absorbed power-law (line with low-energy cutoff at $\sim 20$ keV)
with associated reflection component (line showing an edge at 4.44 keV) and neutral iron line 
(observed EW=1.3 keV, rest-frame EW=1.9 keV).}
\end{figure}

One possible explanation is in terms of a cluster hard tail (or a simple power-law emission) 
as detected by BeppoSAX in Coma and A2199 (Fusco-Femiano et al. 1999,
Kaastra et al. 1999) superimposed on the cluster thermal component. However,
 we regard this possibility as highly unlikely
for the following reasons: a) the hard tail photon index required by the data
is much flatter ($\Gamma$ $\le$ 0.9) than expected, b) the luminosity of the
power law component dominates the 10-100 keV emission, while typically hard tails
are responsible for 10-15$\%$ of the high energy output and c) 
this model would not explain the 6.4 keV iron line emission. 

We therefore, prefer the alternative explanation of an 
AGN deeply buried within a source immersed in the IC plasma.
This requires that a high column density material (N$_{\rm H}$ $\ge$
10$^{24}$ cm$^{-2}$) absorbs the X-ray emission from the AGN, to avoid spoiling the 
good fit of the low energy data with the free-free model. Such high column densities
are quite common in
Seyfert 2 galaxies, as typically evidenced by high energy observations
(Bassani et al. 1999a).  We have therefore added an
absorbed power-law plus a narrow gaussian line to take
into account the obscured AGN emission. This results in a satisfactory fit
($\chi^{2}/\nu$ =47.2/63) and returns a column density of 
$N_{\rm H}\simeq 6.7^{+32.7}_{-1.3}$ $\times$ 10$^{24}$ cm$^{-2}$ 
for a fixed photon index of 1.9. However, the EW of the 6.4  keV iron line
would turn out to have an unrealistically high value ($>>$ 10 keV), i.e the line 
must be extremely intense to survive the heavy absorption and to emerge above the cluster 
emission. It is therefore not compatible with 
pure transmission throughout the measured column density (Leahy \& Creighton 1993).
If present, the line must be produced in a different way, e.g. by reflection 
in the absorbing torus or in the surrounding material.

Consequently the data have been fitted using a more complex model
assuming thermal emission from the cluster plus AGN emission,
the latter including an absorbed power-law, to reproduce the X-ray flux transmitted
through the absorbing torus, and a reflection spectrum with associated neutral 
iron line. Note that modelling
the data with a pure reflection component cannot account for all of the 
high energy excess, so that both AGN components are required by our observations. 

Obviously, given the limited statistics of our data, various parameters of this
complex description (the primary power-law photon index [$\Gamma$=1.9], 
line energy [6.4 keV] and width [0] and cluster metal abundance [$Z=0.4$]) have to be fixed.
In this case, good  fits ($\chi^{2}/\nu$ =48-51/62) are obtained  if
N$_{\rm H}$ is $\ge$ 5 $\times$ 10$^{24}$ cm$^{-2}$, 
and if the fraction R of the nuclear flux reflected over all directions is
in the range R=0.15-0.3. For this choice of the parameters, the EW of the neutral
Fe line is $\sim$ 1-2 keV and the cluster temperature is 5.5 keV, 
quite consistent with the assumed metal abundance (see illustration in Figs. 3
and 4).

\section{Discussion }

IRAS 09104+4109 clearly shows a very complex phenomenology when observed
in X-rays. Particularly informative about the dual nature of the source
(including IC plasma emission at 0.1-7 keV and AGN emission above 7 keV)
are our new BeppoSAX observations. In particular, the PDS data 
demonstrate the energetic dominance of AGN emission at the high energies, and
add to independent evidence based on optical line spectroscopy and
polarization measurements.

The best-fit parameters found for our cluster and AGN emissions are not
"per se" unusual. In particular, the inferred AGN spectrum is similar 
to what is found in other highly absorbed Seyfert 2 galaxies, such as Mkn3, 
NGC4945, NGC6240 and the Circinus galaxy (Bassani et al. 1999b; Vignati 1999).
Rather, what is unusual is to find the combined cluster/AGN phenomenon in a
single source, and, for a Type-2 AGN like this, the absorption-corrected luminosity:
0.8 and 1.7 $\times$ 10$^{46}$  erg s$^{-1}$ in the 2-10 and 10-100 keV
band respectively. These luminosities are well within the range of QSO
observed values.  

In spite of their limited statistics, and thanks to the very wide 
dynamic range in photon energy,  BeppoSAX data are able to constrain the complex
physical situation in the source. To get physically consistent solutions,
our fitting procedure required two emission components in the AGN, one in
transmission through the torus, the other in reflection to explain the neutral
Fe line. The torus is expected to absorb low energy photons along
obscured directions and, if $N_{\rm H}$ is large, to produce a reflection bump
along unobscured directions. Indeed,
while substantial absorption is confirmed, in particular, by the detection of
the silicate 10 $\mu$m absorption feature by Taniguchi et al. (1997),
anisotropic emission and scattering are clearly consistent with a variety of
observations in the radio (the usual two-lobe structure) and optical (the highly
polarized ionization cones detected by the HST imaging polarimetry, see Hines et
al. 1999).

\begin{figure}
\psfig{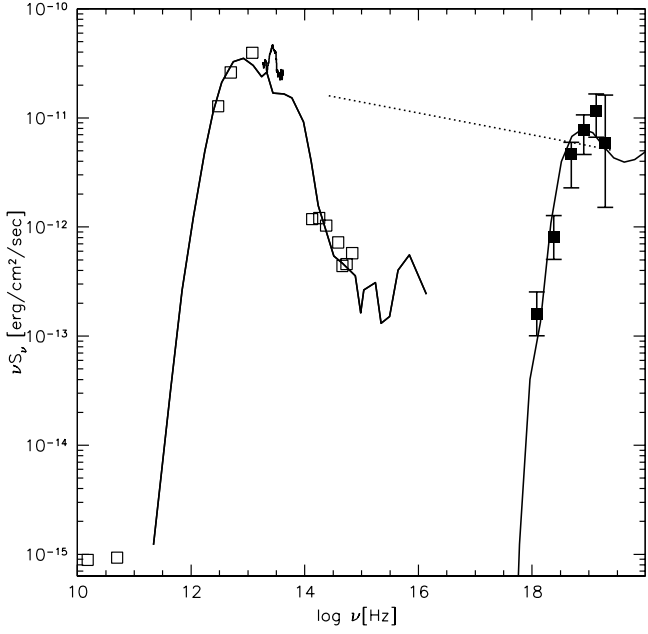}
\caption{The observed broad-band spectrum of the type-2 quasar IRAS 09104+4109 over 10 decades 
in photon energy. The cluster emission dominating the low X-ray energies has been
omitted.
The IR and hard X-ray spectra are fitted with the same self-consistent AGN model as
described in Sect. 6.
This includes a dusty torus structure around the central AGN and responsible for the
IR emission (Granato et al. 1996), with the associated gas assumed to photoelectrically
absorb and Compton scatter the X-ray photons. 
The dotted line shows a schematic fit to the average X-ray to IR spectrum of type-1 AGNs 
(Barcons et al. 1995) adopted by us to describe the AGN primary continuum. 
The continuous lines show the effect of the processing by the surrounding dusty torus.
Datapoints are from Taniguchi et al.
(1997) for the IR and from BeppoSAX observations for the X-rays (two data points at 5 
and 10 keV [$log\nu$=18.08 and $log\nu$=18.36] come from subtraction of the cluster 
contribution).
}
\end{figure}

A rough match of the 2-10 keV luminosity with the OIII (corrected for reddening
in the Narrow Line Region, Bassani et al. 1999a) and with IR emissions
indicates L$_{\rm X}$/L$_{\rm OIII}$ and L$_{\rm X}$/L$_{\rm Fir}$ ratios of 18
and 0.3, i.e. fairly typical for Type-1 AGN (Bassani et al. 1999a, Mulchaey et
al. 1994).

We have performed a detailed comparison of the observed hard X-ray and IR
continua with predictions of radiative transfer codes (Granato, in preparation)
accounting for (dust and photoelectric) absorption and scattering, including
the anisotropy and incoherence of Compton scattering
(particularly relevant for the X-rays). The codes have been tested against
and found to be in agreeement with Monte Carlo simulations. 

We have then adopted for the gaseous torus in IRAS 09104+4109 the same 
geometry needed to reproduce the IR continuum (Granato et al. 1996) and
the ionization cones (Hines et al. 1999). Consistent with the results of our spectral
analysis in Sect. 5, this geometry predicts a thick torus with opening
angles of $\sim 30$ degrees observed with an angle of the symmetry plane with
the line-of-sight of  $\sim 45$ degrees. Then the predicted X-ray flux within
the two frequencies $\nu_1$ and $\nu_2$ is given by:

\begin{equation}
F(\nu_1,\nu_2)=\frac{1}{4\pi d_{\rm L}^2} L^{\rm QSO}_{\rm 5 keV}
P_{\nu_{1e},\nu_{2e}}(N_{\rm H})
\label{pippo}
\end{equation}

\noindent where $\nu_e=\nu (1+z)$ and $L^{\rm QSO}_{5 keV}$
is the unobscured luminosity for a type-1 quasar, and
$P_{\nu_{1e},\nu_{2e}}(N_{\rm H})$ is the thickness-dependent ratio between
the flux emerging along lines of sight with column density $N_{\rm H}$ and the
flux coming along unobscured directions at 5 keV, 
as derived from the solution of the radiative transfer equation. 

As for the IR emission, the observed monochromatic flux at 60 $\mu$m is given by

\begin{equation}
F_\nu(60 \mu{\rm m})=\frac{L_\nu^{\rm QSO}(60 \mu{\rm m} /[1+z])}{4\pi d_{\rm L}^2}
\, (1+z) \, p_{60/[1+z]}(\tau_e,\Theta)
\label{pluto}
\end{equation}

\noindent where $p_{60/[1+z]}(\tau_e,\Theta)$ is the ratio in the rest-frame
between the monochromatic flux at $60/[1+z]\simeq 40\ \, \mu$m 
emitted by the dusty torus with optical thickness $\tau_e$ along the
line of sight (defined by the polar angle $\Theta$) and
the flux emitted along a typical unobscured direction.
Combining the two above equations we get finally:

\begin{equation}
F(\nu_1,\nu_2)=
\frac{F_\nu(60 \mu{\rm m}) \, R \, P_{\nu_{1e},\nu_{2e}}(N_{\rm H})}
{p_{60/[1+z]}(\tau_e,\Theta) \, (1+z)}
\label{topolino}
\end{equation}
\noindent where $R=L^{\rm QSO}_\nu({\rm 5 keV})/L^{\rm QSO}_\nu(60/[1+z]
\mu{\rm m})$.

Barcons et al.\ (1995) find  an average $f_\nu({\rm 5
keV})/f_\nu(12 \mu{\rm m})= 1.4^{+1.1}_{-0.4} \times 10^{-6}$ for a
sample of 54 Seyferts 1 (error corresponding to 95\% confidence).
Adopting the mean IR SED for UVX quasars as in Elvis et al. (1994),
this translates into an X-ray to IR ratio of $R \simeq
1.3 \times 10^{-6}$. Assuming $N_{\rm H}=6\ 10^{24}$, $\tau_e=N_{\rm H}/1.1\ 10^{21}$,
$\Theta=45$ degrees,
and for the observed monochromatic flux $F_\nu(60 \mu{\rm m})=550$  mJy,
we obtain the hard X-ray spectrum reported in Fig. 5. The expected flux at 20-100
keV turns out to be $F(\nu_1,\nu_2)\simeq 8\ 10^{-12}$ erg/cm$^2$/sec, whereas
the observed is $\simeq 10^{-11}\pm 0.3$ erg/cm$^2$/sec.

%
%
%
%
%
%

Altogether, within the observed dispersion of the ratio of X-ray to IR flux for
quasars, and within the uncertainties in the torus model geometry, the
high-energy flux of IRAS 09104+4109 detected by BeppoSAX is consistent with
emission by the same quasar which produces, via the reprocessing by the
circumnuclear dust, the whole far-IR spectrum.

\section{Conclusions}

BeppoSAX observations of IRAS 09104+4109 provide 
evidence for the existence of a buried AGN and 
indicate that at least some objects classified as HyLIRGs can indeed 
harbour highly luminous X-ray sources, the long-sought high luminosity/redshift
analogues of Seyfert2 galaxies.

In principle, this result could have implications for the long standing
debate about the nature of the primary power source in ultraluminous IR galaxies
and the related question of origin of the cosmological IR and X-ray backgrounds.

However, we have to consider the rather unique nature of this source, where the
simbiosis of a bright IR quasar with a rich IC plasma including a massive
cooling-flow is emphasized by the present BeppoSAX observations.

If we consider the rarity of luminous type-2 QSOs (mostly found in IRAS
surveys, e.g. probably IRAS F10214, IRAS F15307 and IRAS 23060, very rarely detected in
X-ray surveys, e.g. RXJ13434+0001), then it is tempting to
relate the peculiarity of IRAS 09104+4109 with the huge cooling-flow of $\sim
1000\ M_\odot$ discovered by ROSAT HRI (Fabian \& Crawford 1995). This
condition of extremely fast mass accumulation could counter-balance the enormous
radiation pressure by the quasar (which would otherwise quickly get rid of the obscuring
envelope), hence favouring the persistence of a thick medium around it.

%

\begin{acknowledgements}
This research has made use of SAXDAS linearized 
and cleaned event files produced at the BeppoSAX Science Data Centre. 
Research partially supported by Italian Space Agency (ASI).
\end{acknowledgements}

\end{document}